\documentclass{elsart}
\usepackage{epsfig}
\usepackage{graphicx}
\usepackage{amsmath,amssymb}
\begin{document}

\newcommand{\EP}{\mbox{e$^+$}}
\newcommand{\EM}{\mbox{e$^-$}}
\newcommand{\EPEM}{\mbox{e$^+$e$^-$}}
\newcommand{\EMEM}{\mbox{e$^-$e$^-$}}
\newcommand{\EE}{\mbox{ee}}
\newcommand{\EL}{\mbox{e}}
\newcommand{\G}{\mbox{$\gamma$}}
\newcommand{\Z}{\mbox{$Z$}}
\newcommand{\GG}{\mbox{$\gamma\gamma$}}
\newcommand{\GE}{\mbox{$\gamma$e}}
\newcommand{\TEV}{\mbox{\,TeV}}
\newcommand{\GEV}{\mbox{\,GeV}}
\newcommand{\LGG}{\mbox{$L_{\gamma\gamma}$}}
\newcommand{\LGE}{\mbox{$L_{\GE}$}}
\newcommand{\LEE}{\mbox{$L_{\EE}$}}
\newcommand{\WGG}{\mbox{$W_{\gamma\gamma}$}}
\newcommand{\EV}{\mbox{\,eV}}
\newcommand{\CM}{\mbox{\,cm}}
\newcommand{\MM}{\mbox{\,mm}}
\newcommand{\NM}{\mbox{\,nm}}
\newcommand{\MKM}{\mbox{\,$\mu$m}}
\newcommand{\SEC}{\mbox{\,s}}
\newcommand{\CMS}{\mbox{\,cm$^{-2}$s$^{-1}$}}
\newcommand{\MRAD}{\mbox{\,mrad}}
\newcommand{\IND}{\hspace*{\parindent}}
\newcommand{\E}{\mbox{$\epsilon$}}
\newcommand{\EN}{\mbox{$\epsilon_n$}}
\newcommand{\EI}{\mbox{$\epsilon_i$}}
\newcommand{\ENI}{\mbox{$\epsilon_{ni}$}}
\newcommand{\ENX}{\mbox{$\epsilon_{nx}$}}
\newcommand{\ENY}{\mbox{$\epsilon_{ny}$}}
\newcommand{\EX}{\mbox{$\epsilon_x$}}
\newcommand{\EY}{\mbox{$\epsilon_y$}}
\newcommand{\BI}{\mbox{$\beta_i$}}
\newcommand{\BX}{\mbox{$\beta_x$}}
\newcommand{\BY}{\mbox{$\beta_y$}}
\newcommand{\SX}{\mbox{$\sigma_x$}}
\newcommand{\SY}{\mbox{$\sigma_y$}}
\newcommand{\SZ}{\mbox{$\sigma_z$}}
\newcommand{\SI}{\mbox{$\sigma_i$}}
\newcommand{\SIP}{\mbox{$\sigma_i^{\prime}$}}
\newcommand{\fb}{\mbox{\,fb}}
\newcommand{\lsim}{\raisebox{-0.07cm}{$\,
\stackrel{<}{{\scriptstyle\sim}}\, $}}
\newcommand{\gsim}{\raisebox{-0.07cm}{$\,
\stackrel{>}{{\scriptstyle\sim}}\, $}}

\begin{frontmatter}
\title{
\hspace*{9cm} {\large\rm Budker INP 2002-47} \\[1cm]
%\hspace*{13cm} {\large July 6, 2002} \\
Interaction region for $\gamma \gamma, \gamma e$ collisions
  at linear colliders} 
\thanks{Talk at the VIII Intern. Conference on
  Instrumentation for Colliding Beam Physics, Novosibirsk, Russia,
  Feb.~28-March~6, 2002} 
 
\author{Valery Telnov} 
\address{Institute of Nuclear Physics, 630090 Novosibirsk, Russia \\
e-mail: telnov@inp.nsk.su}
\date{}

\begin{abstract}
  Photon colliders (\GG, \GE) are based on backward Compton scattering
  of laser light off the high energy electrons in linear colliders.
  All projects of linear colliders include this option. In this paper
  physics motivation, possible parameters and some interaction region
  aspects of photon colliders are discussed.

\end{abstract}
\end{frontmatter}

\vspace{-0.7cm}
\section{ Introduction}
\vspace{-0.5cm}
Linear colliders, LC, at the center of mass energy from 100 GeV up to
several TeV will be one of the central instruments in experimental
high energy physics in the next 2 to 3 decades.  Four projects are
being developed: NLC~\cite{NLC}, TESLA~\cite{TESLAall},
JLC~\cite{JLC}, and CLIC~\cite{CLIC}.  TESLA team has published in
March 2001 the Technical Design of the superconductiing linear
collider on the energy 90--800 GeV~\cite{TESLAall}.

  The unique feature of the $e^+e^-$ Linear Colliders is the
 possibility to construct on its basis a Photon Collider using the
 process of the Compton backscattering of laser light off the high
 energy electrons~\cite{GKST83,GKST84,TEL90,TEL95}. 
 This option is considered now for all linear colliders projects: NLC
 \cite{NLC}; TESLA \cite{TESLAgg,telnov,TESLATDR}; JLC
 \cite{JLCgg,takahashi,JLC}; CLIC \cite{burkhardt}. 
 
 In the following we discuss the physics programme, possible laser and
 optical schemes, the expected \GG\ and \GE\ luminosities and some
 other interaction region aspects.  Examples are given mostly for the
 photon collider at TESLA, other projects are also discussed 
  where they have principle differences.
 
\begin{figure}[!htb]
 \centering 
  \vspace*{0.0cm}
\epsfig{file=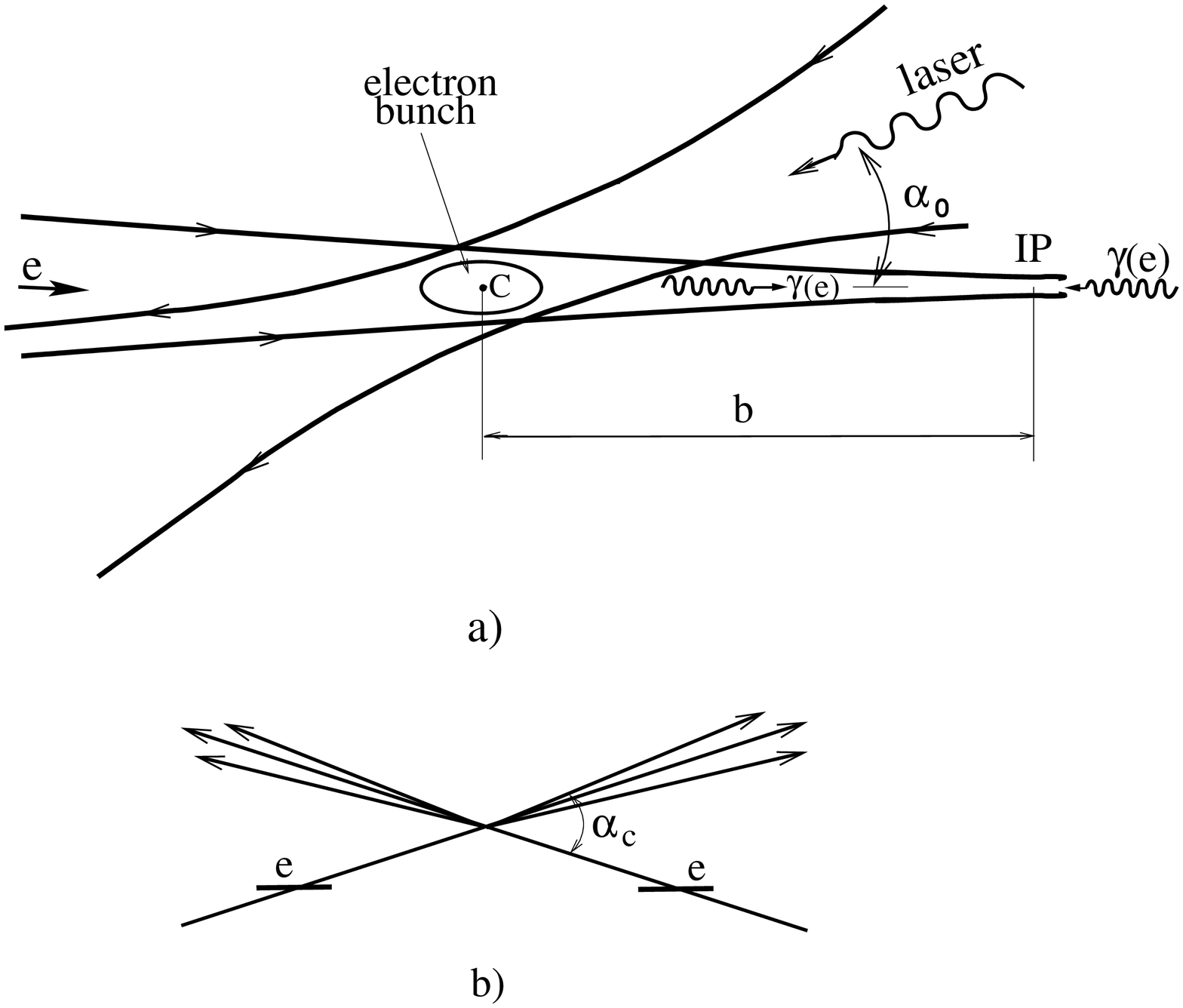,width=8.cm,angle=0}
 \vspace*{-0.4cm}
\caption{Scheme of \GG, \GE\ collider.}  
\vspace*{-0.3cm}
\label{ggcol}
\end{figure}

 The basic scheme of the Photon Collider is shown in
 Fig.~\ref{ggcol}. Discussion of the photon collider
 scheme and basic principles can be found
 elsewhere~\cite{GKST83,GKST84,TEL90,TEL95,TESLAgg,telnov,TESLATDR}. 

 Let us remind main
 features of photon colliders. The maximum energy of the scattered photons is~\cite{GKST83}
\begin{equation}
\omega_m=\frac{x}{x+1}E_0; \;\;\;\;
x \approx \frac{4E_0\omega_0}{m^2c^4}
 \simeq 15.3\left[\frac{E_0}{\TEV}\right]\left[\frac{\omega_0}{\EV}\right]=
 19\left[\frac{E_0}{\TEV}\right]
\left[\frac{\MKM}{\lambda}\right],
\label{eq1}
\end{equation}
where $E_0$ is the electron beam energy and $\omega_0$ the energy of
the laser photon. For example, for $E_0 =250 \GEV$, $\omega_0
=1.17\EV$ ($\lambda=1.06\MKM$) (powerful solid state lasers) we
obtain $x=4.5$ and $\omega_m = 0.82 E_0 = 205 \GEV$ (it will be
somewhat lower due to nonlinear effects in Compton scattering.

The value $x\approx 4.8$ is a
good choice for photon colliders, because for $x > 4.8$ the
produced high energy photons create QED \EPEM\ pairs in collision with
the laser photons, and as result the \GG\ luminosity is 
reduced~\cite{GKST83,TEL90,TEL95}.  Hence, the maximum center of mass
system energy in \GG\ collisions is about 80\%, and in \GE\ 
collisions 90\% of that in \EPEM\ collisions. If for some study lower
photon energies are needed, one can use the same laser and decrease
the electron beam energy. The same laser with $\lambda \approx 1 \MKM$ 
can be used for all TESLA energies. At $2E_0 = 800\GEV$ the parameter 
$x\approx 7$, which is larger than 4.8. But  nonlinear effects at the 
conversion region effectively increase the threshold for \EPEM\ production,
so that \EPEM\ production is significantly reduced~\cite{TESLATDR}.
 
The luminosity distribution in \GG\ collisions has a high energy
peak and a low energy part.  The peak has a width at half
maximum of about 15\%. The photons in the peak can have a high degree
of circular polarization. This peak region is the most useful for
experimentation. When comparing event rates in \GG\ and \EPEM\ 
collisions usually we use the value of the \GG\ luminosity in this peak
region $z>0.8z_m$ where $z=\WGG/2E_0$.

The energy spectrum of high energy photons becomes most peaked if the
initial electrons are longitudinally polarized and the laser photons
are circularly polarized. This gives almost a factor of 3--4 increase
of the luminosity in the high energy peak.  The average degree of the
circular polarization of the photons within the high-energy peak
amounts to 90--95\%.  The sign of the polarization can easily be
changed by changing the signs of electron and laser polarizations.

The \GG\ luminosity in the high energy luminosity peak for
TESLA is just proportional to the geometric luminosity $L_{geom}$ of
the electron beams: $\LGG(z>0.8z_m) \approx 0.1 L_{geom}$.  The
latter can be made larger for \GG\ collisions than the \EPEM\
luminosity because beamstrahlung and beam repulsion are absent for
photon beams.  It is achieved using beams with smallest possible
emittances and stronger beam focusing in the horizontal plane (in
\EPEM\ collisions beams should be flat due to beamstrahlung).
For current TESLA parameters ~\cite{telnov,TESLATDR} 
(see details in Section 3) 
\vspace{-0mm}
\begin{equation}
 \LGG (z>0.8z_m) \approx \frac{1}{3} L_{\EPEM}.
  \label{eq2}
\vspace{-0mm}
\end{equation}
Approximately similar luminosities are expected in \GE\ collisions.
Typical cross sections  in \GG\ collisions are
about one order of magnitude higher than those in \EPEM\ collisions
therefore the number of interesting  events at photon colliders will
be similar of even higher than in \EPEM\ collisions.

\vspace{-0.5cm}
\section{Physics motivation}
\vspace{-0.5cm}
The physics potential of the Photon Collider is very
rich and complements in an essential way the physics program of the
\EPEM\ mode. A few examples: 
\begin{itemize}
\item In \GG\ collisions, resonances with $C=+$ are produced as single
  resonances. The effective cross section of the $h^0$ production at
  photon colliders is even higher that in \EPEM\ 
  collisions~\cite{ee97}. One of the most important examples is the
  Higgs boson.  The precise knowledge of its two--photon width is of
  particular importance.  It is sensitive to heavy virtual charged
  particles.  Supersymmetry predicts three neutral Higgs bosons.
  Photon colliders can produce the heavy neutral Higgs bosons with
  masses about 1.5 times higher than in \EPEM\ collisions at the same
  collider (because  heavy Higgs bosons $A^0$ and
  $H^0$ have almost equal masses and in \EPEM\ collisions are produced
  in pairs).
\item 
  A \GG\ collider can produce  pairs of any charged particles
  (charged Higgs, supersymmetric particles etc.) with a cross section
  about one order of magnitude higher than those in \EPEM\ collisions
  (see graphs elsewhere~\cite{TEL95,TESLAgg,TESLATDR}.
  Moreover, the cross sections depend in a different form on various
  physical parameters.  The polarization of the photon beams
  and the large  cross sections allow to obtain valuable
  information on these particles and their interactions.
\item 
  At a \GE\ collider charged particles can be produced with masses higher
  than in pair production of  \EPEM\ collisions (like a new $W^{\prime}$ boson
  and a neutrino or a supersymmetric scalar electron plus a neutralino).
\item  
  Photon colliders offer unique possibilities for measuring the \GG\ fusion of
  hadrons for probing the hadronic structure of the photon.
\end{itemize} 
 A short list of physics processes for the photon collider is presented in
Table~\ref{processes}~\cite{summary}.  More detail consideration of the
physics program at photon colliders can be found
elsewhere~\cite{BERK,ee97,GG2000,summary,TESLATDR,JLC}.

\begin{table}[!hbtp]
\caption{Gold-plated processes at photon colliders}
\vspace{3mm}
{\renewcommand{\arraystretch}{1.0} \small
\begin{center}
\begin{tabular}{ l  c } 
\hline
$\quad$ {\bf Reaction} & {\bf Remarks} \\
\hline\hline
$\GG\to h^0 \to b\bar{b},\GG$ & $M_{h^0}<160$ \GEV  \\
$\GG\to h^0 \to WW(WW^*)$    & $140<M_{h^0}<190\,\GEV$ \\
$\GG\to h^0 \to ZZ(ZZ^*)$      & $180<M_{h^0}<350\,\GEV$ \\
\hline
$\GG\to H,A \to b\bar{b} $  &
 MSSM heavy Higgs \\
$\GG\to \tilde{f}\bar{\tilde{f}},\
\tilde{\chi}^+_i\tilde{\chi}^-_i,\ H^+H^-$ & SUSY particles \\ 
$\GG\to S[\tilde{t}\bar{\tilde{t}}]$ & 
$\tilde{t}\bar{\tilde{t}}$ stoponium  \\
$\GE \to \tilde{e}^- \tilde{\chi}_1^0$ & $M_{{\tilde{e}^-}} < 
0.9\times 2E_0 - M_{{\tilde{\chi}_1^0}}$\\
\hline
$\GG\to W^+W^-$ & anom. inter., extra dim. \\
$\GE^-\to W^-\nu_{e}$ & anom. $W$ couplings \\
$\GG\to WW+WW(ZZ)$ & strong $WW$ scattering\\
\hline
$\GG\to t\bar{t}$ & anom. $t$-quark interact. \\
$\GE^-\to \bar t b \nu_e$ & anom. $W tb$ coupling \\
\hline
$\GG\to$ hadrons & total \GG\ cross section \\
$\GE^-\to e^- X$ and $\nu_{e}X$ &  struct. functions \\ 
$\gamma g\to q\bar{q},\ c\bar{c}$ & gluon distr. in the \G\ \\
$\GG\to J/\psi\, J/\psi $ & QCD Pomeron \\
\hline
\end{tabular}
\end{center}
}
\label{processes}
\end{table}

\vspace{-2mm}
\section{Luminosities at photon colliders}
\vspace{-5mm}

\subsection{The collision scheme, crab--crossing}
\vspace{-5mm}

The basic scheme for photon colliders is shown in
Fig.~\ref{ggcol}.  The distance between the conversion
point (CP) and the IP, $b$, is chosen from the relation $b \approx
\gamma \sigma_y$, so that the size of the photon beam at the IP has
equal contributions from the electron beam size and the angular spread
from Compton scattering. At TESLA $\sigma_y \approx 4\NM$ gives $b
\approx 2\MM$ at $2E_0=500\GEV$. Larger $b$ values lead to a decrease
of the \GG\ luminosity. For smaller $b$ values the high energy
luminosity increases only a little while the low--energy photons
give a larger contribution to the luminosity (which is not useful for
the experiment but causes additional backgrounds).

The removal of the disrupted beams can best be done using the
crab-crossing scheme~ Fig.~\ref{ggcol}, which is
foreseen in the NLC and JLC projects for \EPEM\ collisions.  Due to
the collision angle the outgoing disrupted beams travel outside the
final quads.  The value of the crab--crossing angle is determined by
the disruption angles and the final quad design (diameter of the quad
and its distance from the IP).  Simulation shows that for TESLA the maximum
disruption angle is about 10--12 mrad. Above this angle the total
energy of particles in disrupted beams is smaller than from 
unremovable \EPEM\ pairs backgrounds.  In the present TESLA design
$\alpha_c = 34\MRAD$.

\vspace{-5mm}
\subsection{Collision effects, ultimate luminosities} 
\vspace{-5mm}

At first sight, one may think that there are no collision
effects in \GG\ and \GE\ collisions because at least one of the beams
is neutral. This is not correct because during the beam collision
electrons and photons are influenced by the field of the opposite
electron beam, which leads to the following effects~\cite{TEL90,TEL95}:

\vspace{-2mm}

$\bullet$ in \GG: conversion of photons into \EPEM\ pairs 
         (coherent pair creation);

\vspace{-2mm}

$\bullet$ in \GE: coherent pair creation; beamstrahlung;
   beam displacement.

\vspace{-2mm}

Beam collision effects in \EPEM\ and \GG, \GE\ 
collisions are different.  In particular, in \GG\ collisions there are
no beamstrahlung or beam instabilities which limit the horizontal beam
size in \EPEM\ collisions on the level 550 (350) nm for TESLA
(NLC/JLC). Therefore, it was of interest to study limitations of the
luminosity at photon colliders due to beam collision effects.  The
simulation, which include all collision effects, was done for the
TESLA beams and the horizontal size of the electron beams was varied.
The results are the following (see graphs in
\cite{Tfrei,telnov,TESLATDR}): all curves for the \GG\ luminosity
follow their natural behavior (as without collision effects):
$L\propto 1/\sigma_x$ at least down to $\sigma_x = 10\NM$ (smaller
values were not considered because too small horizontal sizes may
introduce problems with the crab--crossing scheme). The \GG\ 
luminosity (in the high energy part) can reach the value
$10^{35}$\CMS. 
Note that while in \EPEM\ collisions $\sigma_x\approx 500\NM$,
in \GG\ collisions the attainable $\sigma_x$ with the planned injector
(damping ring) is about $100\NM$.  In \GE\ collisions the luminosity
at small $\sigma_x$ is lower than follows from the geometric scaling
due to beamstrahlung and displacement of the electron beam during the
beam collision. Its maximum value (high energy part) is about
(2--3)$\cdot 10^{34}$ \CMS. Corresponding numbers and curves for
NLC/JLC beam parameters can be found in \cite{Tfrei}.

So, we can conclude that for \GG\ collisions at c.m.s. energies
below about one TeV one can use beams with $\sigma_x$ much smaller than in
\EPEM\ collisions.  As a result, the \GG\ luminosity in the high
energy peak can be, in principle, several times higher than the \EPEM\ 
luminosity. Contrary to \EPEM\ collisions, where the luminosity is
determined by unremovable beam-beam collision effects, at photon
colliders the achievable \GG\ luminosity is limited only by technical 
problems in obtaining beams with small transverse sizes.

\vspace{-5mm}
\subsection{$\gamma\gamma$ and $\gamma$e luminosities at TESLA}
\vspace{-5mm}

Attainable \GG\ luminosity depends strongly on the emittances of the
electron beams. There are two methods of production low--emittance
electron beams: damping rings and low--emittance RF--photo--guns
(without damping rings).  The second option is promising, but at the
moment there are no such photo--guns producing polarized electron
beams.  Polarization of electron beams is very desirable for photon
colliders.  So, there is only one choice now --- damping rings.
Especially for a photon collider the possibility of decreasing the beam
emittances at the TESLA damping ring has been studied~\cite{Decking}
and it was found that the horizontal emittance can be reduced down to 
$\ENX\ = 2.5\times 10^{-6}$ m.   

The luminosity also depends on the $\beta$--functions at the
interaction points: $L \propto 1/\sqrt{\beta_x \beta_y}$.  The
vertical $\beta_y$ is usually chosen close to the bunch length
$\sigma_z$ (this is the design for \EPEM\ collisions and can also be
realized for \GG\ collisions).  Some questions remain about the
minimum horizontal $\beta$--function. For \EPEM\ collisions, $\beta_x
\approx 15\MM$ which is larger than the bunch length $\sigma_z =
0.3\MM$, because beams in \EPEM\ collisions must be flat to reduce
beamstrahlung. In \GG\ collisions, $\beta_x$ could be about $1\MM$ (or
even somewhat smaller).  There are two fundamental limitations: the
beam length and the Oide effects (radiation in final quads). The
latter is not important for the beam parameters considered. There is
also a certain problem with the angular spread of the synchrotron
radiation emitted in the final quads. But, for the photon collider the
crab--crossing scheme will be used and in this case there is
sufficient clearance for the removal of the disrupted beams and
synchrotron radiation.  Studies of the existing
scheme for the TESLA final focus have shown~\cite{Walker} that
chromo--geometric aberrations dominate at $\beta \leq
6\MM$. Fortunately, a new scheme for the final focus system proposed
at SLAC~\cite{Sery} (see also graphs in \cite{TESLATDR}) allows to obtain
$\beta_x \approx 1.5\MM$ with small aberrations and further
optimization is possible.  

The parameters of the photon collider at TESLA for $2E_0=$ 200, 500
 and 800 GeV are presented in Table~\ref{sumtable}. For
 comparison the \EPEM\ luminosity at TESLA is also included.  It is
 assumed that the electron beams have 85\% longitudinal polarization
 and that the laser photons have 100\% circular polarization. The
 thickness of the laser target is one Compton scattering length for
 $2E_0=$ 500 and 800 GeV and 1.35 scattering length for $2E_0=$ 200
 GeV, so that $k^2 \approx$ 0.4 and 0.55, respectively ($k$ is the $e
 \to \gamma$ conversion coefficient).  The laser wave length is 1.06
 \MKM\ for all energies.  The distance between conversion and
 interaction points is $b=\gamma\sigma_y$ for $2E_0=500$ and 800 GeV
 and $b=2\gamma\sigma_y$ for $2E_0=200$ GeV. Simulation results
 presented below include nonlinear effects in the Compton
 scattering~\cite{Galynskii}. Corresponding parameters $\xi^2 =
 0.15, 0.2,0.4$ for $2E_0=200,500,800$ GeV, respectively.
\begin{table}[!tp]
\vspace{-0mm}
\caption{Parameters of  the photon collider at TESLA.}
\vspace{4mm}
{\renewcommand{\arraystretch}{1.} \setlength{\tabcolsep}{0.8mm} \small
\begin{center}
\begin{tabular}{l c c c} \hline
$2E_0$,GeV & 200 & 500 & 800   \\ \hline
$\lambda_L$ [\MKM]/$x $& 1.06/1.8 & 1.06/4.5 & 1.06/7.2 \\
$t_{L}/\lambda_{scat}$& 1.35 & 1 &1 \\
$N/10^{10}$& 2 & 2 & 2  \\  
$\sigma_{z}$ [mm]& 0.3 & 0.3 & 0.3  \\  
$f_{rep}\times n_b$ [kHz]& 14.1 & 14.1 & 14.1  \\
$\gamma \epsilon_{x/y}/10^{-6}$ [m$\cdot$rad] & 2.5/0.03 & 2.5/0.03 & 
2.5/0.03 \\
$\beta_{x/y}$ [mm] at IP& 1.5/0.3 & 1.5/0.3 & 1.5/0.3 \\
$\sigma_{x/y}$ [nm] & 140/6.8 & 88/4.3 & 69/3.4  \\  
b [mm] & 2.6 & 2.1 & 2.7 \\ \hline
\LEE(geom) [$10^{34}$\,\CMS] & 4.8 & 12 &  19 \\ 
$\LEE (z>0.65)  $ & 0.03 & 0.07 & 0.095 \\  \hline
 $W_{\GG,\,max}$ (GeV)& $122$ & $390$ & $670$ \\  
$\LGG (z>0.8z_{m,\GG\ })  $ [$10^{34}$] & 0.43 & 1.1 &  1.7  \\ \hline
 $W_{\GE,\,max}$ (GeV)& $156$ & $440$ & $732$ \\
$\LGE (z>0.8z_{m,\GE\ })$ [$10^{34}$]  & 0.36 & 0.94 & 1.3 \\  \hline \hline
$L_{\EPEM}$ [$ 10^{34}$\,\CMS]  & 1.3 & 3.4 & 5.8 \\

%$\theta_{max}$, mrad & 10 & 9 & 16 \\ 
\end{tabular}
\end{center}
}
\label{sumtable}
\vspace{0.4cm}
\end{table}   
From Table~\ref{sumtable} one can see that for the same energy 
$\LGG(z>0.8z_m) \approx (1/3) L_{\EPEM}.$ 
\begin{figure}[!bt]
\centering
\vspace*{-0.cm}

\epsfig{file=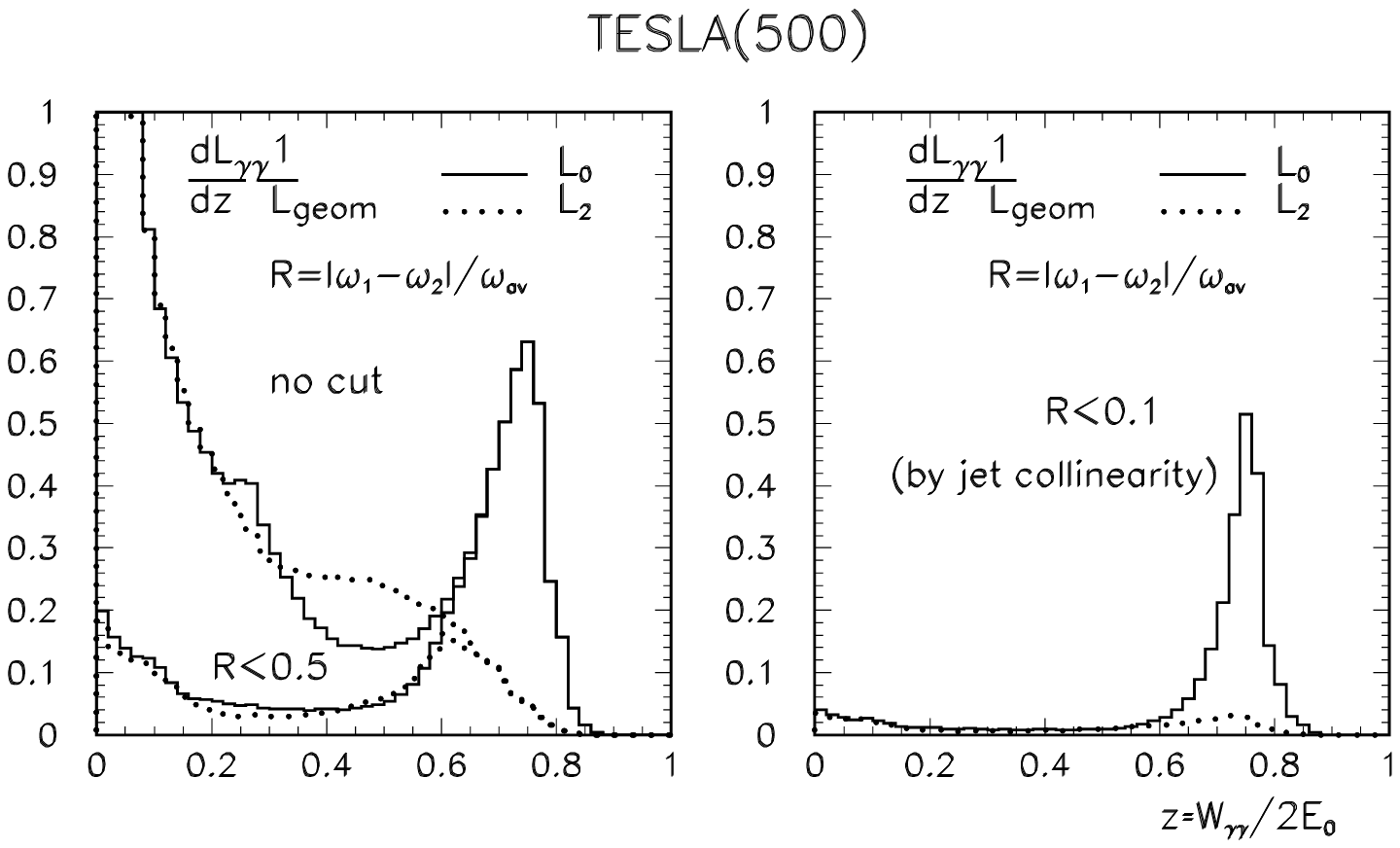,height=7.5cm,width=14.cm,angle=0}

\vspace{-2cm}

\epsfig{file=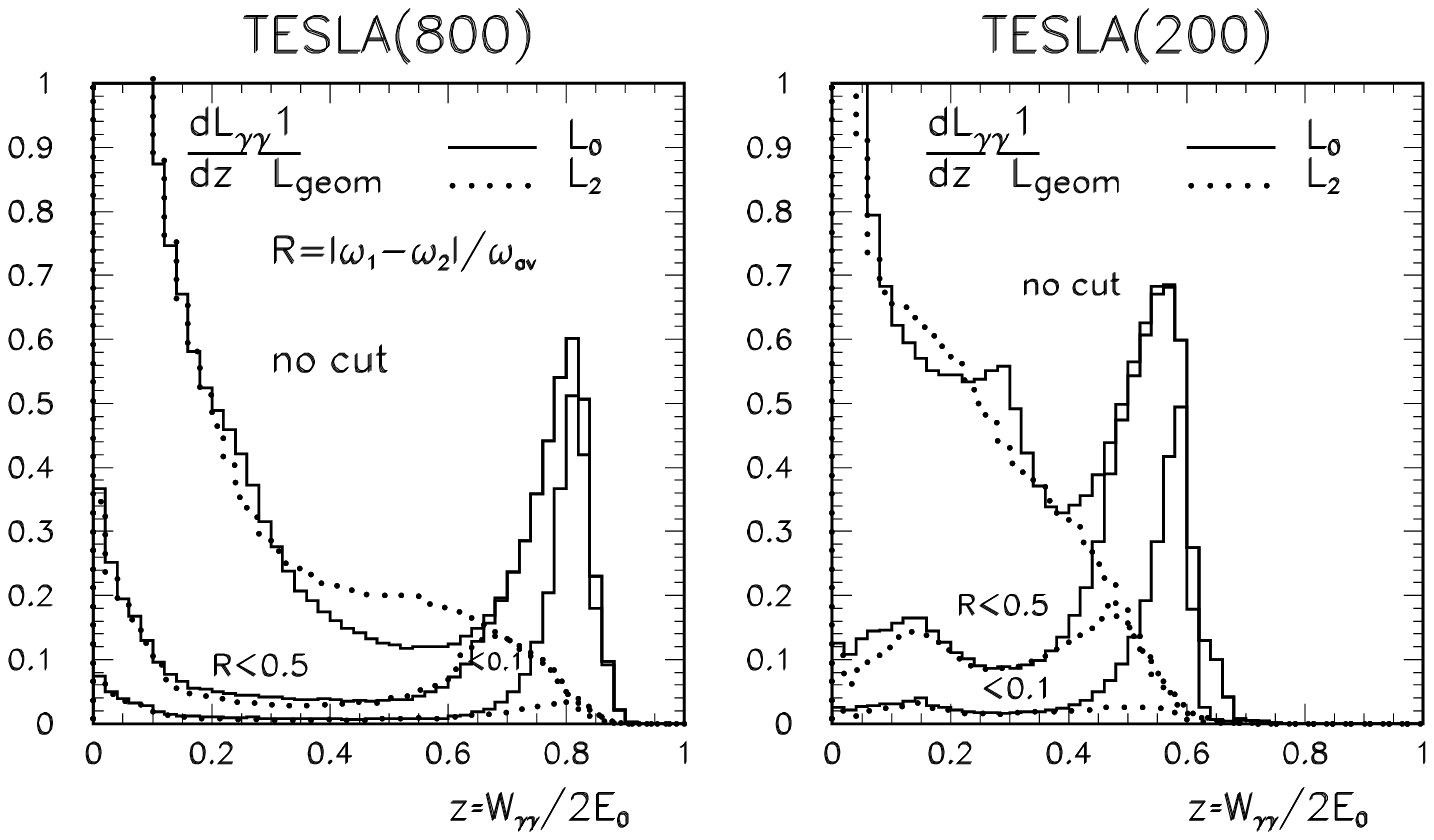,height=7.5cm,width=14.cm,angle=0}

\vspace{-1.2cm}

\caption{\GG\ luminosity spectra at TESLA. 
Solid line for total helicity of the two photons 0 and dotted line
  for total helicity 2. Dashed curves: luminosities  with cuts on
longitudinal momentum. See the text. }
\label{Ldist250}
\vspace*{-0.cm}
\end{figure}
\begin{figure}[!htb]
\centering
\vspace*{-0.7cm}
\hspace{-0.0cm}   \epsfig{file=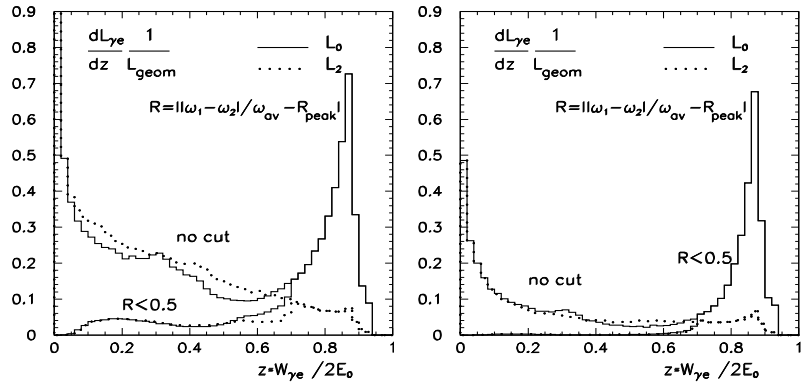,height=7.cm, 
width=13.2cm,angle=0}
\vspace{-1.cm}
\caption{Normalized \GE\ luminosity spectra at TESLA(500). Left:
the photon collider is optimized for \GG\ collisions; right: one electron 
beam is converted and the distance between IP and CP is 1.05\CM,
 5 times larger than for the left figure.}
\vspace{0.3cm}
\label{Ldist250e}
\end{figure}
Simultaneously with \GG\ collisions there are also \GE\ collisions
with somewhat lower luminosity, so one can study both types of
collisions simultaneously.  Residual electron-electron luminosity is
very small due to the beam repulsion.

The normalized \GG\ luminosity spectra for $2E_0 = 500$ GeV and 800,
200 GeV are shown in Fig.~\ref{Ldist250}. The luminosity spectra are
decomposed in two parts: with the total helicity 0 and 2.
Fig.~\ref{Ldist250} shows also luminosity  spectra with
additional cuts on the longitudinal momentum of the produced system,
which suppress the low energy luminosity to a low level.  In the
case of two jets one can restrict the longitudinal momentum using
the acollinearity angle between jets ($H\to b\bar b, \tau\tau$, etc.).

For the Higgs the production rate is proportional to $dL_0/dW_{\GG}$
at $W_{\GG}=M_h$.  For the case considered, $M_h \approx
120\GEV$, and $x=1.8$, $dL_0/dW_{\GG} = 1.7\times 10^{32}\CMS/\GEV.$
Note that the corresponding peak luminosity for NLC at the time of
Snowmass2001 was 7 times smaller\cite{NLC} and more recent revised number is
 2.8 times smaller~\cite{asner}. In turn, TESLA also has large resources
for further improvement of the luminosity. It will be more clear after
detailed optimization of the final focus  system. 

The normalized \GE\ luminosity spectra for $2E_0=500$ \GEV\ and
parameters from Table~\ref{sumtable} are shown in
Fig.~\ref{Ldist250e}-left.  For dedicated $\GE$ experiments one can
convert only one electron beam, increase the distance between the
conversion and the interaction points and obtain the 
\GE\ luminosity spectrum with suppressed low energy part,
Fig.~\ref{Ldist250e}-right.

The \GG\ luminosity distributions, including their polarization
characteristics, can be measured using processes $\gamma\gamma\to
l^+l^-$ ($l=e$, $\mu$)~\cite{TEL93,TESLATDR}. For measurement of \GE\ 
one can use the processes $\gamma$e$\to
\gamma$e~\cite{GKST83,TESLATDR}, $\GE\to \EL\ \EPEM
(\EL\ \mu^+\mu^-)$~\cite{GKST83} and $\GE\to eZ^0$.

\vspace{-5mm}
\section{Lasers-Optics} 
\vspace{-5mm}

A key element of photon colliders is a powerful laser system which is
used for the $e \to\gamma$ conversion. The required parameters and possible
schemes of lasers are discussed below.  

\vspace{-5mm}
\subsection{Requirements for lasers}
\vspace{-5mm}

 There are two main parameters characterizing Compton scattering: $x$
defined by Eq.\ref{eq1} and  $ \xi^2 =
e^2\overline{B^2}\hbar^2/(m^2c^2\omega_0^2) = 2 n_{\gamma} r_e^2
\lambda/\alpha$ characterizing nonlinear effects in Compton
scattering.  At $\xi^2 \ll 1$ the electron scatters on one laser
photon, at $\xi^2 \ll 1$ -- on several.  With grows of $\xi^2$ the
spectrum becomes wider and is shifted to lower
energies~\cite{Galynskii,TESLATDR}, therefore  it is preferable to have
$\xi^2 << 1$. However, decrease of $\xi^2$ in the conversion leads
inevitably to the increase of the required laser flash energy. For TESLA
project we assume $\xi^2 = 0.15, 0.3. 0.4$ for $2E_0 = 200, 500, 800$ GeV
when change of the luminosity spectra is still acceptable.

Another (now ``good'') consequence of the nonlinear effects is the
shift of the threshold for \EPEM\ pair creation in collision of laser
and high energy photons. Instead of $x \approx
4.8$~\cite{GKST83,TEL90,TEL95} for the linear Compton it is $x_{th}=
4.8(1+\xi^2)$. For the maximum TESLA energy $2E_0=800$ GeV and laser
wavelength 1.06\MKM\ $x \approx 7.2$.  For this energy we assume
$\xi^2 =0.4$ and $x_{th} \approx 6.7$.  As result, pair creation is
practically absent and we can use the same laser in the whole TESLA
energy region.

   For optimization of conversion efficiency we have performed
simulation asuming that the final focusing mirrors are situated outside
the electron beams (without holes in mirrors  for electron beams). The tilte of
beams due to the crab crossing angle  was taken into
account. The result of the simulation~\cite{telnov} of the conversion 
probability for the electron bunch length $\sigma_z= 0.3\MM$, 
$\lambda=1.06\MKM$, $x=4.8$ as a function of the
Rayleigh length $Z_R$ for various flash energies and values of the
parameter $\xi^2$ can be found elsewhere~\cite{telnov,TESLATDR}.

 In order to obtain a conversion
probability of $k\approx 63\%$  at all TESLA energies a laser with the
following parameters is required:
\begin{center}
{\renewcommand{\arraystretch}{1.0}
\begin{tabular}{ll}
Flash energy & A $\approx 5$ J \\
Rayleigh length & $Z_R \sim 0.4$ mm \\
Duration & $\tau(rms) \approx 1.5$ ps \\
Repetition rate & TESLA collision rate, $\nu \approx 14$ kHz \\
Average power & $P \approx 140$ kW (for one pass collisions) \\
Wavelength &   $\lambda \approx 1$ \MKM\ (for all energies).
\end{tabular}
}
\end{center}   

For NLC/JLC colliders the electron bunch is shorter by a factor of
two. Calculations show that for similar assumptions (laser optics is
outside the electron beams) the required flash energy is 3.3
J and it is about 1.75 J for the case head-on collisions (mirrors with the
holes) which is considered for these projects.

\begin{figure}[!htb]
\centering
\vspace*{-0.cm} 
\mbox{\epsfysize=5cm\epsffile{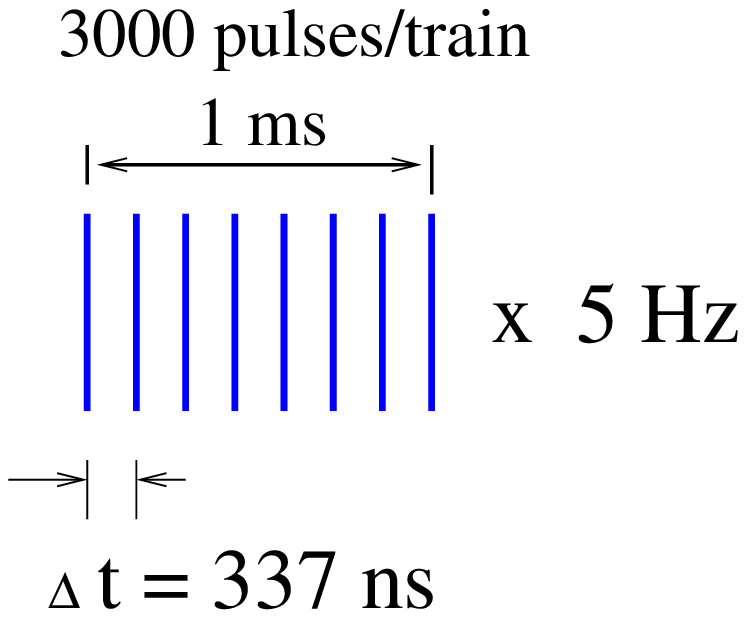}}
\mbox{\vspace{-0cm}\epsfysize=5.cm\epsffile{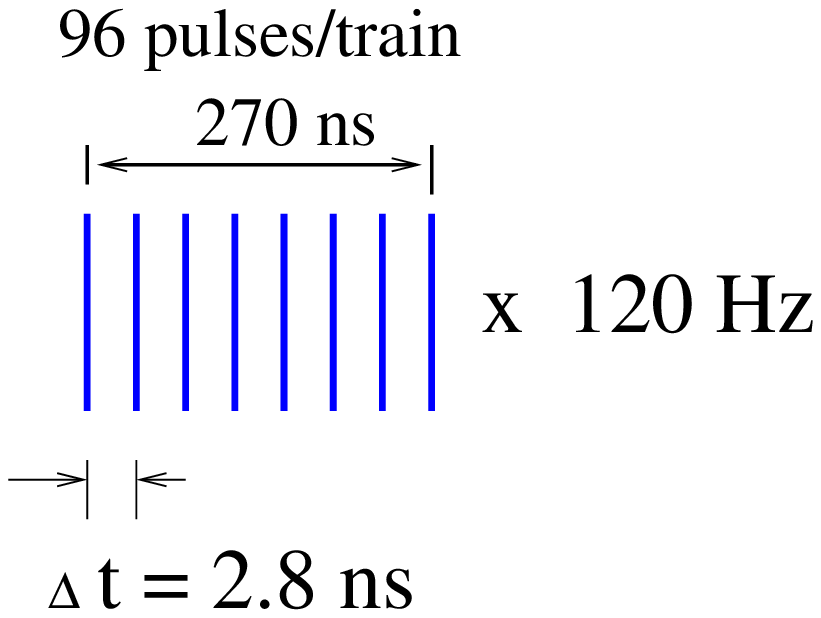}}
\caption{Structure of electron beams at TESLA(left) and NLC/JLC(right).}
\label{trains}
\vspace{-0mm}
\end{figure}

Main difference for lasers between TESLA and NLC projects is connected
with structures of electron bunch trains, see Fig.\ref{trains}. At
superconducting TESLA the distance between bunches is 100 times larger
which leads to different possible approaches for laser schemes.

\vspace{-2mm}
\subsection{Laser schemes}
\vspace{-5mm}

All parameters for lasers are reasonable for exception of the
repetition rate (average power).  To overcome the ``repetition rate''
problem it is quite natural to consider a laser system where one laser
bunch is used for the $e \to \gamma$ conversion many times.  Two ways
of multiple use of one laser pulse are considered for the Photon
Collider at TESLA: an optical storage ring and an external optical
cavity.

 In the first scheme~\cite{telnov,TESLATDR},
each laser bunch is used for the $e \to
\gamma$ conversion about 12 times.  The laser pulse is sent to the
interaction region where it is trapped in an optical storage ring.
This can be done using Pockels cells (P), thin film polarizers (TFP)
and 1/4-wavelength plates ($\lambda/4$).  The maximum number of cycles
is determined by reflection coefficients of mirrors and attenuation in
the Pockels cell.

\begin{figure}[!htb]
\centering
\vspace*{0.cm} 
 \epsfig{file=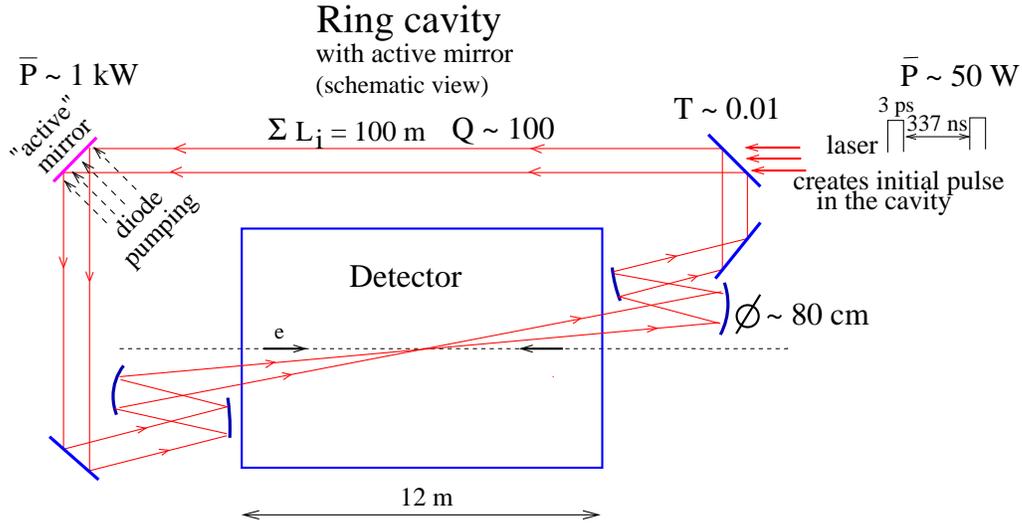,width=14.cm,angle=0} 
\caption{Ring type ``external'' optical cavity for the photon collider
  at TESLA (only optics for conversion of one electron beam is shown).}
\label{cavity}
\vspace*{.4cm} 
\end{figure}

In the second scheme~\cite{Tfrei,telnov,Will,TESLATDR},
an ``external'' optical cavity is used. Using a train of low energy
laser pulses one can create in the ``external'' cavity (with one
mirror having small transmission) an optical pulse with an energy
higher than in a laser pulse by a factor Q (cavity quality factor) and
this pulse collides with electron bunches many times.  As result the
required laser power can be lower than in the one-pass case by factor
of 50--100. Recently, I.Will~\cite{WillKrakow} has suggested to place
the final focusing mirrors outside the detector and added an
``active'' mirror to the optical cavity which should compensate
attenuation of the laser pulse. Initial 5 J pulse in the cavity is
created using a train of specially prepared pulses from the external
laser (about 200 pulses, 0.05 J, 3 ps long, with frequencies matched
to the cavity). Then this pulse is circulating in the optical ring
cavity and losses of its intensity are compensated by an active
mirror.  Schematically this approach is illustrated in
Fig.\ref{cavity}. The optical scheme includes also (not shown) movable
mirrors, adaptive optics, fast feedback. There are many questions to
this scheme which need further study.

For NLC/JLC the train is short and the distance between electron
bunches is only 2.8 ns therefore exploiting of the optical cavity is
not effective.  Small distance between electron bunches makes also a
problem for combining pulses from several lasers to one train using
Pockels cells. Current solution is the following. The laser system is
based on about Mercury diode pumped lasers ($\lambda \approx 1 \MKM$
(Yb:S-FAP)) developed for the fusion program.  The laser produces
100-200 J, several ns long pulses with 10 Hz rep. rate. Each pulse is
split using semitransparent mirrors into 96 pulses with 2.8 ns delays
and which then are compressed down to 2 ps using grating pairs. In
this way one Mercury pulse is sufficient for production of one $96
\times 1.5$ J pulse train for NLC photon collider. Twelve such lasers
working sequentially are needed to provide 120 Hz rep. rate. The
estimated cost of such laser ``plant'' according to the report at
Snowmass2001 is of the order of 200 M\$~\cite{NLC}.
 
This approach does not look optimum because only one Mercury laser
of twelve is used in a given moment. This means, for example, that one
need 12 independent diode-pumping systems which are very costly. It
seems more reasonable to use about 5 special $\sim$30 J lasers which work
in parallel with 120 Hz rep. rate. Each 30 J pulse is split in the
same way as in \cite{NLC} to twenty 1.5 J pulses, ten with vertical and
ten with horizontal direction of linear polarization. Subtrains with
equal polarization from all lasers are combined using Pockels cells
and two long trains with perpendicular polarizations can be joined to
one train using a thin film polarizer. Note that in this case the
required switching time for Pockels cells is 33 ns which looks
possible. The number of lasers and corresponding switching times can
be optimized.
 
\begin{figure}[!htb]
\centering
\vspace*{0.cm} 
 \epsfig{file=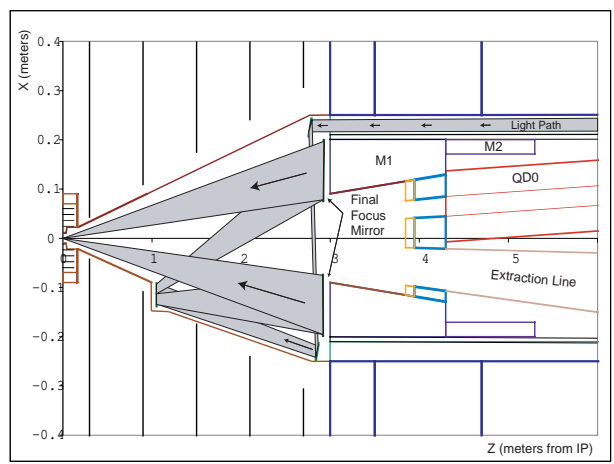,width=8.0cm,angle=0} 
\caption{Laser optics at the NLC interaction region.}
\label{fgron}
\vspace*{-0.cm} 
\end{figure}

The optical system inside the detector for one NLC approach is
shown in Fig.\ref{fgron}~\cite{NLC}. It is one pass laser scheme. The final
focusing mirrors of 38 cm diameter are situated at the distance 4 m
from the IP. The mirror has a hole with 15 cm diameter for incoming beams
and outgoing disrupted beams. 

Development of laser technologies is being driven by several large
programs, such as inertial fusion.  This is a fortunate situation for
photon colliders as we may benefit from the laser technology
developments of the last $10$--$15$ years which cost hundreds M\$ per
year.  In the last decade the technique of short pulse powerful solid
state lasers made an impressive step and has reached petawatt
($10^{15}$) power levels and few femtosecond durations.  Obtaining few
joule pulses of picosecond duration is not a problem using modern
laser techniques.  The wave length of the most powerful lasers about
1\MKM\ which is just optimum for the TESLA Photon Collider.

\vspace{-3mm}
\section{Other experimental issues}
\vspace{-5mm}
\subsection{Backgrounds}
\vspace{-5mm}
Photon colliders have one very specific background problem. High
energy photon beams are produced just inside the detector. After the
$e \to \gamma$ conversion the electron beams have a very broad
spectrum, $E \approx (0.02-1)E_0$, and an angular spread of about 10
mrad which is due to deflection of the low energy particles by the
opposing electron beam. In the case of head-on beam collisions large
fraction of the beam would hit the final quadrupoles and produce
enormous backgrounds in the detector. This problem is solved by using
crab-crossing scheme of beam collisions~\cite{palmer,TEL90}.  The
maximum disruption angle for TESLA photon collider is about 12
mrad~\cite{TESLATDR}. In the present design the crab-crossing angle in
the photon collider IP is 34 mrad. These values put restriction on
possible quadrupole design.

Other backgrounds are \EPEM\ pairs and hadrons produced in beams
collisions. Consideration of these and some other backgrounds can be
found elsewhere \cite{TESLAgg,sit1,TESLATDR}. QED background at photon
colliders are very similar to \EPEM\ collsions (or even smaller due to
the large crab crossing angle and exit hole for disrupted beams.
 Strong longitudinal detector magnetic field confines
low energy \EPEM\ pairs near the collider axis and considerably reduce
backgrounds in the vertex detector. 

Some problem presents a hadronic background. The
effective average rate of the reaction $\GG\to hadrons$ is about
one event per bunch collisions. It can lead to some worsening of the
resolution for jets at low angles~\cite{TESLATDR}.

\vspace{-5mm}   
\section{Conclusion}
\vspace{-5mm}

As the ECFA Panel in Europe and the Snowmass Study and the HEPAP panel
in US have recommended the linear collider on the energy about 500 GeV
as the next large HEP project (Asian physicists have also similar
plans), it is very likely that in about one decade physicists will get
a new very powerful instrument for study of matter: \EPEM, \GG, \GE,
\EMEM\ collider. 

  The photon collider with \GG, \GE\ collisions is very
natural and technically feasible option for linear colliders. The
physics potential of the photon collider is very rich and give
information complementary to that in \EPEM\ collisions.  
The next step in development of the photon collider,
designing the second interaction region and detector needs active
participation of experimentalists, detector physicists.

\vspace{0mm}   
The work has been partially supported by the INTAS 00-00679 grant.

\end{document}